\documentstyle[twoside,fleqn,espcrc2a,epsfig]{article}

\def \pbarp {p{\overline p}}

\newcommand{\AmS}{{\protect\the\textfont2
  A\kern-.1667em\lower.5ex\hbox{M}\kern-.125emS}}

\title{Comparison of Three-jet Events in $\pbarp$ Collisions at
$\sqrt{s}=1.8$ TeV to Predictions
from a Next-to-leading Order QCD Calculation}

\author{Sally Seidel \\
	Department of Physics and Astronomy,
	University of New Mexico, Albuquerque, NM 87131 USA \\ 
	\ \\
	{\sl For the CDF Collaboration}
       }
       
\begin{document}

\begin{abstract}
The properties of three-jet events with total transverse 
energy greater than 320~GeV and individual jet energy greater than
20~GeV 
have been analyzed and compared to absolute predictions 
from a next-to-leading order (NLO) perturbative QCD calculation.  
These data, of integrated luminosity 
86~pb$^{-1}$, were recorded by the CDF Experiment for $\pbarp$ collisions at 
$\sqrt{s}=1.8$ TeV.
This study 
tests a model of higher order QCD processes that result in gluon emission
and can be used to estimate the magnitude of the contribution of processes
higher than NLO.  The total cross section is measured to be 
$466\pm 3({\rm stat.})^{+207}_{-70}({\rm syst.})$~pb.
The differential cross section is furthermore measured for all
kinematically accessible regions of the Dalitz plane, including those
for which the theoretical prediction is unreliable.  While the measured
cross section is consistent with the theoretical prediction in magnitude,
the two differ somewhat in shape in the Dalitz plane.
\end{abstract}

\maketitle

In perturbative QCD, hard scattering of the constituent partons in the 
proton and antiproton results in events with large total transverse 
energy, $\sum{E_{\rm T}}$.  
Outgoing scattered partons hadronize and may be detected as hadronic jets.  
Three-jet events can be produced when a hard gluon is radiated from any of 
the initial, intermediate, or final state partons in an event with two 
primary outgoing partons.

We analyze here some properties of the cross section for
three-jet event production in
proton-antiproton collisions at the Fermilab Tevatron Collider at
center-of-mass energy 1.8~TeV.  
The data, which were recorded by the
Collider Detector at Fermilab (CDF)~\cite{kn:Det},
are compared with predictions by the first complete
next-to-leading order (NLO) QCD generator, Trirad~\cite{kn:Kilgore},
for hadronic three-jet
production at hadron colliders.  
We compare the measured and predicted absolute
cross sections to test our understanding of the higher order
QCD processes that result in gluon emission and to estimate the magnitude
of the contribution of processes higher than NLO.\footnote{While no
quantitative estimate of the contribution of next-to-next-to-leading order
processes to the cross section is available at this time,
considerable progress has recently been made in calculating two
loop $2 \rightarrow 2$
parton processes~\cite{kn:glover}, important groundwork for the future.}
In some kinematical regions, we provide a 
measurement of the cross section where the theoretical prediction is not
reliable; this measurement may be a useful guide for theoretical calculations.
We
also compared the shapes of the measured and predicted cross sections
when normalized, to examine the sensitivity of the cross
section to variations in the value of the strong coupling, $\alpha_{\rm s}$.
The data sample corresponds to an integrated 
luminosity of 86~pb$^{-1}$ collected during the 1994-1995 run (Run 1b).

A previous paper~\cite{kn:Abe} examined a smaller dataset
and was limited to a comparison with leading order theoretical
calculations~\cite{kn:Kunszt}.
A subsequent analysis~\cite{kn:Geer1} compared a larger dataset to
predictions from the HERWIG~\cite{kn:March}
parton shower Monte Carlo program and to the NJETS~\cite{kn:Berends}
leading order $2 \rightarrow N$ parton-level prediction.  The NLO
calculation used here has the benefit of reduced renormalization
scale dependence (and
consequently lower systematic uncertainty) as well as a more
reliable description of multijet production throughout phase space. 
This study expands upon the previous investigations by comparing the
data to absolute cross section predictions.  The measurements presented here
include differential
cross sections that may be useful constraints upon parton distribution 
functions. 

We use a coordinate system with the $z$ axis along the proton beam,
transverse coordinate perpendicular to the beam, azimuthal angle $\phi$,
polar angle $\theta$, and pseudorapidity $\eta=-\ln \tan(\theta/2)$.
The analysis uses the CDF calorimeters~\cite{calor},
 which cover the pseudorapidity
range $|\eta|<4.2$.  The calorimeters are constructed in a tower geometry
and are segmented in depth into electromagnetic and hadronic components.
The calorimeter towers are 0.1 unit wide in $\eta$.  The tower widths
in $\phi$ are $15^\circ$ in the central region and $5^\circ$ for $|\eta|$ 
greater than approximately 1.2.

We begin by considering
events from the data sample selected by the trigger requirement
$\sum{E_{\rm T}} >$~175~GeV.  We refer to this 175 GeV as
$E_{\rm tot}^{\rm thr}$ below.  Event reconstruction uses a cone
algorithm~\cite{kn:Abe} described in more detail below.
The transverse energy  is 
defined as ${E_{\rm T}} \equiv {E} \; {\rm sin} \, \theta$, where 
$E$ is the scalar sum of energy deposited in the calorimeter within a
particular cone and
$\theta$ is the angle between the beam direction in the laboratory frame 
and the cone axis.   All  
calorimeter energy clusters~\cite{kn:Abe} in the event
with $E_{\rm T} >$~10~GeV are summed.
The three leading jets in the laboratory frame are used as the basis of 
transformation into the three-jet rest frame.  In the three-jet rest frame, 
the incoming partons are, by convention~\cite{kn:UA1}, labeled partons
1 and 2, and their momenta
are designated $\vec p_1$ and $\vec p_2$, respectively.  The highest 
energy jets in this frame have energies labeled $E_{\rm 3}$, $E_{\rm 4}$, 
and $E_{\rm 5}$ and are ordered such that
${E_{3}} > {E_{4}} > {E_{5}}$.  The outgoing partons associated 
with these jets are correspondingly labeled partons 3, 4, and 5.

A three-jet system in the massless parton approximation can be uniquely 
described by five independent variables~(see Figure 5 in~\cite{kn:Geer2}).
We use the following:
\begin{enumerate}
\item the invariant mass of the
three-jet 
system, $m_{\rm 3J}$ 
\item the cosine of the angle $\theta_3^*$
between the average
beam direction ($\vec {p}_{\rm AV} \equiv \vec{p}_1 - \vec{p}_2$) 
and parton 3 in the three-jet rest frame:
$$
{\rm cos} \, \theta_{3}^{*} \equiv \frac{\vec {p}_{\rm AV} \cdot 
\vec {p}_3}{ \left| \vec {p}_{\rm AV} \right| 
\left| \vec {p}_3 \right|}$$
\item
the cosine of the angle $\psi^*$
between the plane containing the average
beam direction and parton 3 and the plane 
containing partons 3, 4, and 5 in their center of mass frame:
$${\rm cos} \, \psi^{*} \equiv \frac{\left
( \vec {p}_3 \times \vec {p}_{\rm AV} \right) \cdot 
\left( \vec {p}_4 \times \vec {p}_5 \right)}
{\left| \vec {p}_3 \times \vec {p}_{\rm AV} \right| 
\left| \vec {p}_4 \times \vec {p}_5 \right|}
$$
\item the Dalitz variable $X_3$ (see below) for the 
leading jet, and 
\item the Dalitz variable $X_4$ (see below) for the next-to-leading jet.  
\end{enumerate}

The invariant
$m_{\rm 3J}$ is calculated by sorting jets by their energies in the
laboratory
frame, boosting to the rest frame of those with the three highest
energies, re-sorting jets by energy in that frame, then computing
$m_{\rm 3J}=\sum_{i=3}^5 E_{\rm i}$, where the $E_{\rm i}$ are the energies
of jets
3, 4, and 5 in the rest frame.
We have
investigated
the probability that a jet with energy less than the weakest of the three
jets in the laboratory frame
may have an energy greater than $E_{5}$ in the 3-jet rest frame from which
it is excluded by this algorithm.  The restriction imposed by the cut
on full trigger efficiency (see below) makes this probability negligible.

The Dalitz variables, $X_{\rm i}$, are defined as
${X}_{\rm i} \equiv {2 \cdot {E}_{\rm i}}/{m_{\rm 3J}}, (i=3,4,5)$.
Momentum conservation restricts the ranges of the Dalitz variables to
\begin{eqnarray}
\frac{2}{3}&\leq \; {X}_{3} \; \leq&1, \nonumber \\
\frac{1}{2}&\leq \; {X}_{4} \; \leq&1, \; {\rm and} \nonumber \\
0&\leq \; {X}_{5} \; \leq&\frac{2}{3}. \nonumber
\end{eqnarray}

A set of trigger and offline 
requirements~\cite{kn:Abe2} rejects events associated with 
cosmic rays, beam halo, and calorimeter malfunctions.  Events are required 
to have a reconstructed primary vertex, defined as the vertex with the largest 
$\sum_{i} {P}_{\rm i}$ (where $P_{\rm i}$ is the total momentum of  
particle {\it i} leaving the vertex in the event),
within $|{z}| <$~60~cm. 
Events are defined to have resolved 
multiple interactions if a second vertex with at least ten associated tracks 
is reconstructed in the vertex track detector, and if that vertex is
separated from the primary
one by at least 10~cm.  
Because multiple interactions can change the jet multiplicity in an event,
for example, misidentifying two-jet events as three-jet events, 
events with resolved multiple interactions
are removed.  The number of events with unresolved
multiple interactions in which an additional jet could be misidentified
is estimated to be less than 2\% so no correction
for them is applied.  The resulting effective total 
integrated luminosity of the data sample is 77$\pm$4~pb$^{-1}$, where the
uncertainty reflects both the overall luminosity uncertainty (4.2\%) and the
uncertainty (0.5\%) associated with the removal of resolved multiple
interactions. 

An iterative cone algorithm~\cite{kn:Abe} with 
cone radius ${R} \equiv \sqrt{(\Delta \eta)^{2} + (\Delta \phi)^{2}}=0.7$
is used to identify jets.
Here 
$\Delta \eta = \eta_{2} - \eta_{1}$ and $\Delta \phi = \phi_{2} - \phi_{1}$.
The subscripts 1 and 2 correspond to the axes of the cone and calorimeter
tower, 
respectively. 

Jets that share towers are combined if the total $E_{\rm T}$ of the shared
towers is greater than 75\% of the $E_{\rm T}$ of either jet; otherwise the
towers are assigned to the nearest jet.
Jet energies are corrected~\cite{kn:Abe} for errors in the absolute 
and relative energy scales and for additional energy associated with the 
underlying event.  Since partons that are radiated out of the cone lead to 
the same losses in the theoretical calculation and in the data, out-of-cone 
corrections are not applied. The $E_{\rm T}$ of a jet is calculated from the
reconstructed position of 
the primary event vertex.
All three leading
jets are 
required to have $E_{\rm T}$~$>$~20~GeV and $|\eta|$~$<$~2.0.  Events with 
fewer than three jets are rejected.  To avoid collinear soft gluon
instability in the 
iterative jet clustering algorithm~\cite{kn:Giele}, a cone overlap cut is 
imposed: events are rejected if the distance $\Delta R$ in $\eta$-$\phi$ space 
between the axes of any two of the three leading jets is less than 1.0
(see Figure 5 of \cite{kn:Abe}, which shows that this selection requirement
reduces to approximately zero the probability of the two jets being merged
by the clustering algorithm).  To 
exclude regions in which
the geometrical acceptance~\cite{kn:Geer2} is less than about 95\%, we 
require 
$|{\rm cos}\, \theta_{3}^{*}| < 
\sqrt{1 - \left( E_{\rm tot}^{\rm thr}/{\rm c}^2/{m_{\rm 2J}} 
\right)^{2}}$,
where 
$m_{\rm 2J}$ is the mass of the two leading jets in the 
three-jet system and is defined analogously to $m_{\rm 3J}$. 

\begin{figure}[htb]
\vspace{9pt}
\mbox{\epsfig{file=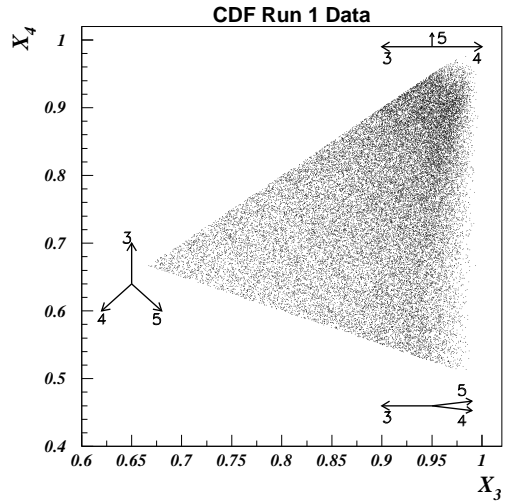,width=3in}}
\caption{The three-jet data, after all selection requirements have been
applied.  The energy correction procedure (see text) has not been applied.
The figures at the corners of the distribution represent typical three-jet
topologies in those regions of the Dalitz plane.}
\label{usec4}
\end{figure}

We require full trigger efficiency, which
occurs when $\sum_{\rm 3~jets}{E_{\rm T}} >$~320~GeV, where the sum is 
over the three highest energy jets in the event with corrected 
$E_{\rm T}$~$>$ 20~GeV~\cite{kn:Abe3}. 
The data are compared to the theoretical prediction by sorting events
into bins of size $0.02 \times 0.02$ in
$X_{\rm 3}$-$X_{\rm 4}$ space, the Dalitz plane.  Figure~\ref{usec4} shows
the Dalitz distribution of data that remain after
all of the selection requirements have been applied.


Before the final binning is done, the data are corrected for the 
effects of the combination of detector resolution and 
energy mismeasurement.   A correction factor is determined for each bin
in the plane as follows.
A sample of events is generated at the parton level with the
HERWIG
Monte Carlo.  The final state partons are hadronized.
The events are then binned in the Dalitz plane.
The same events are next 
passed through the CDF detector simulation and rebinned.  For each bin
the ratio of the number of events after and before detector simulation is
computed.  This ratio (ranging from 0.85 to 1.5)
is the factor subsequently used to
correct the number of events in each data bin. The data are also
corrected for the $z$-vertex cut efficiency and then normalized to the 
effective total luminosity.
 
The principal sources of systematic uncertainty~\cite{kn:syst} on the cross 
section
are those on the absolute and relative 
($\eta$-dependent) jet energy scales.  The uncertainty on absolute
jet energy derives from
the resolution on the calibration of the calorimeter (uncertainty 1.3\%-1.8\%, 
and $E_{\rm T}$-dependent), the uncertainty
associated with choice of
jet fragmentation model (decreasing from 1.7\% to 1.2\% with
increasing $E_{\rm T}$), the uncertainty associated with
calorimeter stability over time (1\%), and the uncertainty on the correction
for the contribution of the underlying event (1 GeV).  The uncertainty
on the relative jet energy scale ranges from 2\% to 6\%.  Uncertainties are 
also associated with the measurement of the effective
total integrated
luminosity (4.2\%) and with the $z$-vertex cut efficiency (2\%).  There is
also an
uncertainty of less than 5\% associated with the implementation of simulated
events in the correction procedure.  The upper (lower) limits on 
these uncertainties are added (subtracted) from the four-momenta of 
the jets in the data sample to obtain the systematic uncertainties on the
cross section associated with each contribution.  The uncertainties are
then combined to produce the total experimental systematic uncertainty.

The Trirad calculation consists of $2 \rightarrow 3$ parton processes
at one loop and $2 \rightarrow 4$ parton processes at tree level.  For
gluons $g$, incoming quarks $q$, and outgoing quarks $Q$ or $Q^\prime$, the
subprocesses involved are $gg \rightarrow ggg$, ${\overline q}q \rightarrow
ggg$, ${\overline q}q \rightarrow {\overline Q}Qg$, and those related
by crossing symmetry, all computed to one loop; and $gg \rightarrow gggg$,
${\overline q}q \rightarrow gggg$, ${\overline q}q \rightarrow
{\overline Q}Qgg$, and ${\overline q}q \rightarrow 
{\overline Q} Q {\overline Q^\prime} Q^\prime$ and the crossed processes
computed at tree level.  The program uses the ``subtraction
improved''~\cite{kn:Giele}
phase space slicing method to implement infrared cancellation.

The cross section is predicted with the CTEQ4M~\cite{cteq4} parton
distribution function (PDF) for each bin in the Dalitz
plane.
The result is multiplied by the effective total integrated luminosity 
of the data to predict a number of events in each bin.  
We restrict the prediction to bins for which $X_3<0.98$; this is necessary
as the perturbative expansion is not reliable
where the three-jet configuration approaches a two-jet configuration.
The 
comparison between the data and the calculation is made for 215 bins.   

\begin{figure}[htb]
\vspace{9pt}
\mbox{\epsfig{file=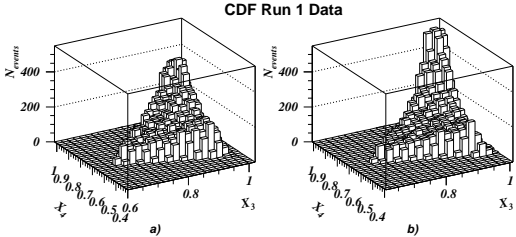,width=3in}}
\caption{The event density in the Dalitz plane for (a) the data and
for (b) the prediction by the NLO
Monte Carlo calculation with CTEQ4M, normalized to luminosity.}
\label{shape}
\end{figure}

Figure~\ref{shape} compares Dalitz distributions of
the data and the absolute theoretical prediction.  The theoretical
distribution is more strongly
peaked---a trend that persists 
in comparisons with all members of the CTEQ4 family\footnote{The CTEQ4
family includes CTEQ4A1, CTEQ4A2, CTEQ4M, CTEQ4A5, and CTEQ4A6, which
differ in the value of $\alpha_{\rm s}$ input to their global fit, and CTEQ4HJ,
for which a higher statistical emphasis was given to the high $E_{\rm T}$
data from CDF.}
of PDFs.  This trend, in
which the edges of the Dalitz plane are more populated by data than by
the prediction, may give some indication of the size of the higher order
contributions to the cross section.


\begin{figure}[htb]
\vspace{9pt}
\mbox{\epsfig{file=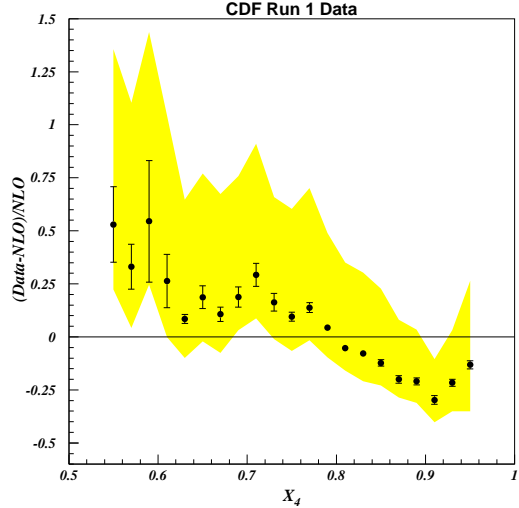,width=3in}}
\caption{The fractional difference between the data and
the theoretical prediction, using CTEQ4M, as a 
function of $X_{4}$, averaged over $X_3$.
The vertical bands show the systematic
uncertainties. The error bars show the statistical uncertainties in cases
where those are larger than the size of the symbol used.
The prediction is normalized to the data to facilitate a comparison
of the shapes of the distributions.}
\label{trendc4}
\end{figure}

\begin{figure}[htb]
\vspace{9pt}
\mbox{\epsfig{file=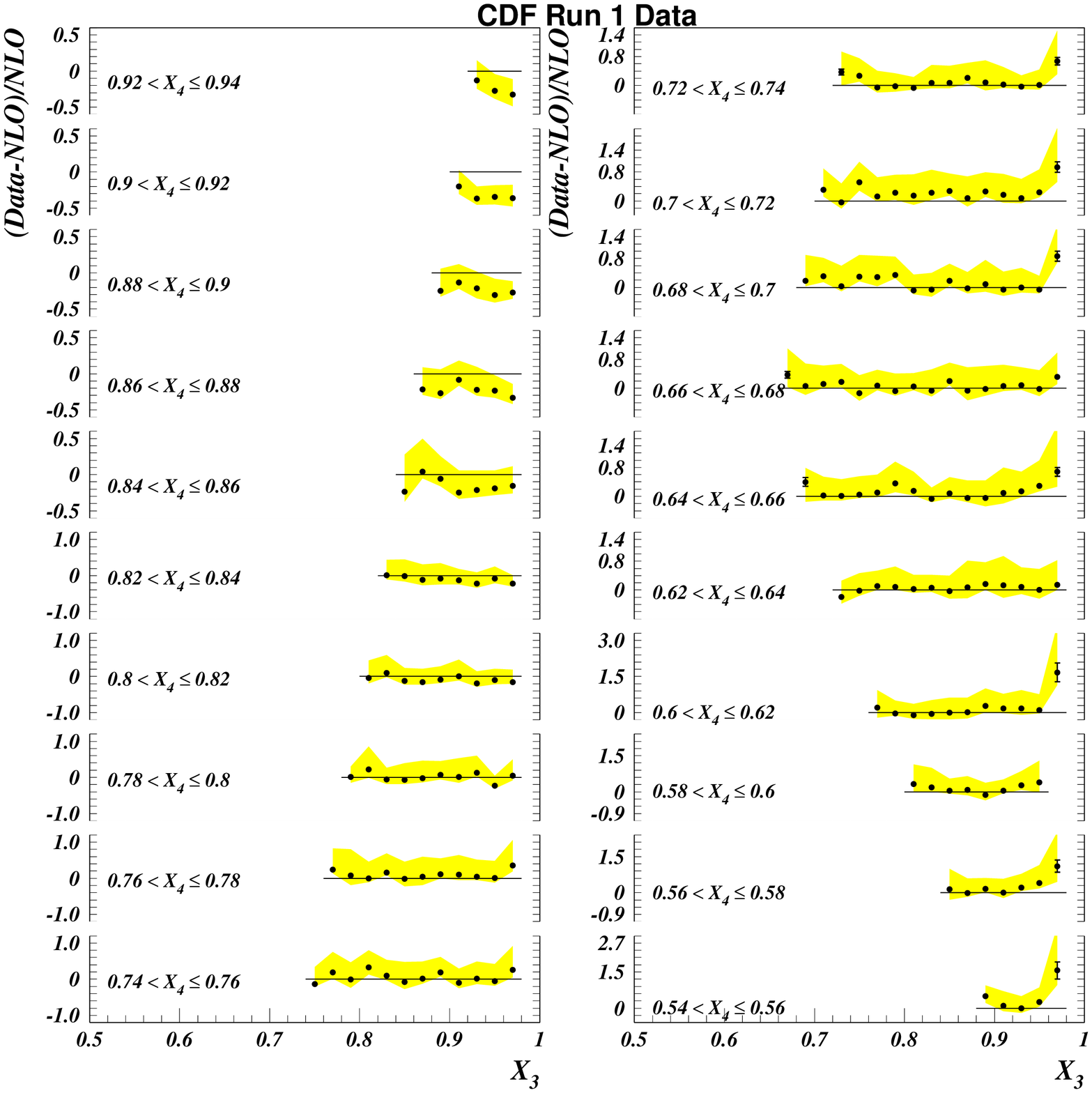,width=3in}}
\caption{The fractional difference between corrected data and the NLO 
prediction, using the CTEQ4M parton distribution function, as a function of 
$X_{3}$ for various $X_{4}$ bins.
Error bars reflect statistical uncertainty for cases in
which it is larger than the size of the symbol used.  Shaded bands indicate
systematic uncertainty.  The prediction is normalized to the luminosity of
the data.}
\label{frac4c7}
\end{figure}

The data and theory are compared in two different ways.  In
Figure~\ref{trendc4}, we
compare the shapes of their Dalitz distributions by normalizing the
data and theory predictions to the same number of events.  In
Figure~\ref{frac4c7},
we normalize theory to the experimental luminosity and compare
the absolute values of the cross sections that are observed and predicted.
In both figures, the prediction is made using the CTEQ4M parton distribution
function, and the difference between observed and predicted number of
events, scaled by the number of predicted events, is computed. 


The theoretical prediction for the cross section, using CTEQ4M and all bins
in the Dalitz plane but those with $X_3>0.98$, is
$473\pm 2({\rm stat.})^{+38}_{-66}({\rm scale})^{+21}_{-28}({\rm PDF})$~pb.
The theoretical uncertainty associated with choice of renormalization and
factorization scales, $\mu_{\rm R}$ and $\mu_{\rm F}$ respectively, 
is estimated by varying the scales, whose default value is $E_{\rm T}$, to
values of $E_{\rm T}/2$ and $2E_{\rm T}$ while keeping
$\mu_{\rm R}=\mu_{\rm F}$.
The theoretical uncertainty associated 
with choice of PDF is estimated from the spread in the predictions generated
with all members of the CTEQ4 family.  The measurement is not sensitive
to the value of $\alpha_{\rm s}$ as is also shown in \cite{kn:CTEQ6}.
The measured cross section, using all bins
in the Dalitz plane but those with $X_3>0.98$, is 
458$\pm$3(stat.)$^{+203}_{-68}$(syst.)~pb.  
This is consistent with the theoretical prediction and with
a previous CDF measurement~\cite{kn:Abe3}
after corrections are made for the efficiencies of 
additional cuts introduced in this analysis. 
The measured cross section, using all bins in the Dalitz plane, is
466$\pm$3(stat.)$^{+207}_{-70}$(syst.)~pb.  

The measurements at high $X_3$
may provide
useful constraints on future theoretical models in that region. 
It appears that up to NLO
the theory predicts more soft radiation than the data have
in the region where the
primary partons are approximately back-to-back.
The data, especially in the region above
$X_3=0.98$ where a perturbative expansion is not reliable, may
be useful input to theoretical models of gluon-emission
processes.

We thank William Kilgore and Walter Giele for providing us with the code 
that computes the next-to-leading order calculation, and for their
guidance concerning its use.  We acknowledge the Center for High
Performance Computing at the University of New Mexico and the University
of Wisconsin Condor Project for providing a combined 26,000 CPU-hours
for the Trirad NLO computations.
We thank the Fermilab staff and the technical staffs of the
participating institutions for their vital contributions.  This work was
supported by the U.S. Department of Energy and National Science Foundation;
the Italian Istituto Nazionale di Fisica Nucleare; the Ministry of Education,
Science, Sports and Culture of Japan; the Natural Sciences and Engineering 
Research Council of Canada; the National Science Council of the Republic of 
China; the Swiss National Science Foundation; the A. P. Sloan Foundation; the
Bundesministerium fuer Bildung und Forschung, Germany; and the Korea Science 
and Engineering Foundation.

\end{document}